\documentclass{article}
\makeatletter
\usepackage[algoruled,boxed,lined]{algorithm2e}
\makeatletter
\g@addto@macro{\@algocf@init}{\SetKwInOut{Parameter}{Parameters}}
\makeatother
\usepackage{amsmath}
\usepackage{amssymb}
\usepackage{graphicx}
\newcommand{\dcocpp}{{\ttfamily dco\kern-.08em{\raisebox{-.1ex}{/}\kern-.15em {c\kern-.03em{\raisebox{-.18ex}{+\kern-.028em{+}}}}}}}

\newcommand{\cpp}{{\ttfamily{C\kern-.03em{\raisebox{-.18ex}{+\kern-.028em{+}}}}}}

\usepackage{hhline}
\usepackage{multirow}

\usepackage{fancyhdr}% http://ctan.org/pkg/fancyhdr
\fancypagestyle{main}{%
  \fancyhf{}
  \rhead{\footnotesize MCMC for Bayesian uncertainty quantification from time-series data}
  \cfoot{\thepage}
}

\usepackage{authblk}

\usepackage[a4paper, total={6in, 10in}]{geometry}

\usepackage{etoolbox}

\makeatletter
\patchcmd{\@maketitle}
  {\addvspace{0.5\baselineskip}\egroup}
  {\addvspace{-2\baselineskip}\egroup}
  {}
  {}
\makeatother

\usepackage{url}

\usepackage[square]{natbib}

\usepackage{floatrow}
\newfloatcommand{capbtabbox}{table}[][\FBwidth]

\begin{document}
\title{\textbf{MCMC for Bayesian uncertainty quantification from time-series data}}
% \author{Philip Maybank\inst{1}\orcidID{0000-0001-8427-2449} \and
% Patrick Peltzer\inst{2} \and
% Uwe Naumann\inst{2} \and
% Ingo Bojak\inst{3}\orcidID{0000-0003-1765-3502}
% }
\author[1]{Philip Maybank}
\author[2]{Patrick Peltzer}
\author[2]{Uwe Naumann}
\author[3]{Ingo Bojak}
\affil[1]{Numerical Algorithms Group Ltd (NAG), \texttt{philip.maybank@nag.co.uk}}
\affil[2]{Software and Tools for Computational Engineering (STCE), RWTH Aachen University,
  \texttt{info@stce.rwth-aachen.de}}
\affil[3]{School of Psychology and Clinical Language Sciences, University of Reading, \texttt{i.bojak@reading.ac.uk}}
\date{}                     %% if you don't need date to appear
\setcounter{Maxaffil}{0}
\renewcommand\Affilfont{\small}
%
% \authorrunning{P. Maybank et al.}
% \institute{Numerical Algorithms Group Ltd (NAG), UK \email{philip.maybank@nag.co.uk} \and
% Software and Tools for Computational Engineering (STCE), RWTH Aachen University, Germany\\
% \email{info@stce.rwth-aachen.de} \and
% School of Psychology and Clinical Language Sciences, University of Reading, UK\\
% \email{i.bojak@reading.ac.uk}}
%
\maketitle              % typeset the header of the contribution
\vspace*{-0.3in}
\begin{abstract}
  Many problems in science and engineering require uncertainty quantification that accounts for observed data.
  For example, in computational neuroscience, Neural Population Models (NPMs) are mechanistic models that
  describe brain physiology in a range of different states.  Within computational neuroscience there is growing
  interest in the inverse problem of inferring NPM parameters from recordings such as the EEG
  (Electroencephalogram).  Uncertainty quantification is essential in this application area in order to infer
  the mechanistic effect of interventions such as anaesthesia.

  This paper presents \cpp{} software for Bayesian uncertainty quantification in the parameters of NPMs from
  approximately stationary data using Markov Chain Monte Carlo (MCMC).  Modern MCMC methods require first order
  (and in some cases higher order) derivatives of the posterior density.  The software presented offers two
  distinct methods of evaluating derivatives: finite differences and exact derivatives obtained through
  Algorithmic Differentiation (AD).  For AD, two different implementations are used: the open source Stan Math
  Library and the commercially licenced \dcocpp{} tool distributed by NAG (Numerical Algorithms Group).

  The use of derivative information in MCMC sampling is demonstrated through a simple example, the
  noise-driven harmonic oscillator.  And different methods for computing derivatives are compared.  The software
  is written in a modular object-oriented way such that it can be extended to derivative based MCMC for other
  scientific domains.
% \keywords{Uncertainty Quantification  \and Algorithmic Differentiation \and Computational Neuroscience}
\end{abstract}
\thispagestyle{fancy}%

\section{Introduction}

Bayesian uncertainty quantification is useful for calibrating physical models to observed data.  As well
as inferring the parameters that produce the best fit between a model and observed data, Bayesian methods can
also identify the range of parameters that are consistent with observations and allow for prior beliefs to be
incorporated into inferences.  This means that not only can predictions be made but also their uncertainty can be
quantified.  In demography, Bayesian analysis is used to forecast the global population
\citep{gerland2014world}.  In defence systems, Bayesian analysis is used to track objects from radar signals
\citep{arulampalam2002tutorial}.  And in computational neuroscience Bayesian analysis is used to compare
different models of brain connectivity and to estimate physiological parameters in mechanistic models
\citep{kiebel2008dynamic}.  Many more examples can be found in the references of
\citep{bishop2006pattern,gelman2013bayesian,lunn2012bugs}.

We focus here on problems which require the use of Markov Chain Monte Carlo (MCMC), a widely applicable
methodology for generating samples approximately drawn from the posterior distribution of model parameters
given observed data.  MCMC is useful for problems where a parametric closed form solution for the posterior
distribution cannot be found.  MCMC became popular in the statistical community with the re-discovery of Gibbs
sampling \citep{smith1993bayesian}, and the development of the BUGS software \citep{lunn2012bugs}.  More
recently it has been found that methods which use derivatives of the posterior distribution with respect to
model parameters, such as the Metropolis Adjusted Langevin Algorithm (MALA) and Hamiltonian Monte Carlo (HMC)
tend to generate samples more efficiently than methods which do not require derivatives \citep{hoffman2014no}.
HMC is used in the popular Stan software \citep{carpenter2017stan}.  From the perspective of a C++ programmer,
the limitations of Stan are as follows: it may take a significant investment of effort to get started.
Either the C++ code has to be translated into the Stan modelling language.  Or, alternatively, C++ code can be
called from Stan, but it may be challenging to (efficiently) obtain the derivatives that Stan needs in order
to sample efficiently.

The software that we present includes (i) our own implementation of a derivative-based MCMC sampler called
simplified manifold MALA (smMALA).  This sampler can be easily be used in conjunction with C++ codes for Bayesian data
analysis, (ii) Stan's MCMC sampler with derivatives computed using \dcocpp{}, an industrial standard tool for
efficient derivative computation.

An alternative approach to the one presented in this paper would be simply to use Stan as a stand-alone tool
without the smMALA sampler and without \dcocpp{}.  Determining the most efficient MCMC sampler for a given
problem is still an active area of research, but at least within computational neuroscience, it has been found the
smMALA performs better than HMC methods on certain problems \citep{sengupta2016gradient}.  Determining the most
appropriate method for computing derivatives will depend on both the user and the problem at hand.  In many
applications Algorithmic Differentiation (AD) is needed for the reasons given in Section \ref{sec:deriv}.  The
Stan Math Library includes an open-source AD tool.  Commercial AD tools such as \dcocpp{} offer a richer set
of features than open-source tools, and these features may be needed in order to optimize derivative
computations.  For example, the computations done using the Eigen linear algebra library \citep{eigenweb} can
be differentiated either using the Stan Math Library or using \dcocpp{} but there are cases where \dcocpp{}
computes derivatives more efficiently than the Stan Math Library \citep{peltzer2019eigen}.  The aim of the
software we present is to offer a range of options that both make it easy to get started and to tune
performance.

\section{Methods for spectral time-series analysis}
\pagestyle{main} % should be second page onwards

\subsection{Whittle likelihood}

The software presented in this paper is targeted at differential equation models with a stable equilibrium
point and stochastic input.  We refer to such models as stable SDEs.  The methods that are implemented
assume that the system is operating in a regime where we can approximate the dynamics through linearization
around the stable fixed point.  If the time-series data is stationary this is a reasonable assumption.  Note
that the underlying model may be capable of operating in nonlinear regimes such as limit cycles or chaos in
addition to approximately linear dynamics.  However, parameter estimation using data in nonlinear regimes
quickly becomes intractable - see Chapter 2 of \citep{maybank2019bayesian}.  The stability and linearity
assumptions are commonly made in the computational neuroscience literature, see for example
\citep{moran2009dynamic}.

In order to simplify the presentation we illustrate the software using a linear state-space model, which is of
the form,
\begin{equation} \label{eq:de}
d\mathbf{X}(t) = A\, \mathbf{X}(t)dt + \mathbf{P}(t),
\end{equation}

\noindent where the term $A\, \mathbf{X}(t)$ represents the deterministic evolution of the system and $\mathbf{P}(t)$
represents the noisy input.  The example we analyze in this paper is the noise-driven harmonic oscillator,
which is a linear state-space model with
\begin{equation}
A =
\begin{pmatrix}
0 & 1 \\
-\omega_0^2 & -2 \zeta \omega_0
\end{pmatrix},
\quad
\mathbf{P}(t) =
\begin{pmatrix}
  0  \\
  dW(t)
\end{pmatrix},
\end{equation}
\noindent and where $dW(t)$ represents a white noise process with variance $\sigma_{in}^2$.  The observations are
modelled as $Y_k = X_0(k\cdot \Delta t) + \epsilon_k$ with $\epsilon_k \sim N(0, \sigma_{obs}^2)$.

Our aim is to infer model parameters ($\omega_0, \zeta, \sigma_{in}$) from time-series data.  This could be
done in the time-domain using a Kalman filter, but it is often more computationally efficient to do inference
in the frequency domain \citep{bojak2005modeling,maybank2017fast}.  In the case where we only have a single
output (indexed by $i$) and a single input (indexed by $j$), we can compute a likelihood of the parameters
$\theta$ in the frequency domain through the following steps.
\begin{enumerate}
\item Compute the (left and right) eigendecomposition of $A$, such that,
\begin{equation}
A \mathcal{R} = \Lambda \mathcal{R}, \quad \mathcal{L} A = \mathcal{L} \Lambda, \quad \mathcal{L} \mathcal{R} = \mathrm{diag}(c) \\
\end{equation}
where $\mathrm{diag}(c)$ is a diagonal matrix, such that $c_i$ is the dot product of the $i$th left
eigenvector with the $i$th right eigenvector.
\item Compute $ij$th element of the transfer matrix for frequencies $\omega_1,\ldots,\omega_K$,
\begin{equation} \label{eq:transfer}
\mathcal{T}(\omega) = \mathcal{R}\ \mathrm{diag}\bigg[ \frac{1}{c_k (i\omega - \lambda_k)} \bigg]\ \mathcal{L}.
\end{equation}
\item Evaluate the spectral density for component $i$ of $\mathbf{X}(t)$, $f_{X_i}(\omega)$, and the spectral
  density for the observed time-series, $f_{Y}(\omega)$,
\begin{align}
  \label{eq:spec_eq}
  f_{X_i}(\omega) &= |\mathcal{T}_{ij}(\omega)|^2\ f_{P_j}(\omega), \\
  f_{Y}(\omega)   &= f_{X_i}(\omega) + \sigma_{obs}^2 \Delta t,
\end{align}
\noindent where $f_{P_j}(\omega)$ is the spectral density for component $j$ of $\mathbf{P}(t)$.
\item Evaluate the Whittle likelihood,
\begin{equation} \label{eq:whittle}
p(y_0,\ldots,y_{n-1}|\theta) = p(S_0,\ldots,S_{n-1}|\theta) \approx \prod_{k=1}^{n/2-1}
\frac{1}{f_Y(\omega_k)} \exp\left[-\frac{S_k}{f_Y(\omega_k)}\right],
\end{equation}
where $\{S_k\}$ is the Discrete Fourier Transform of $\{y_k\}$.  Note that $\theta$ represents a parameter set
(e.g. specific values of $\omega_0, \zeta, \sigma_{in}$) that determines the spectral density.
\end{enumerate}

The matrix $A$ that parameterizes a linear state-space model is typically non-symmetric, which means that
eigenvectors and eigenvalues will be complex-valued.  We use Eigen-AD \citep{peltzer2019eigen}, a fork of the \cpp{} linear
algebra library Eigen \citep{eigenweb}.  Eigen is templated which facilitates the application of AD by overloading tools and Eigen-AD
provides further optimizations for such tools.  The operations above require an AD tool that supports
differentiation of complex variables.  AD of complex variables is considered in \citep{pusch1995ad}.  It is
available from release 3.2.0 of the Stan Math Library and in \dcocpp{} from release 3.4.3.

\subsection{Markov Chain Monte Carlo}
\label{sec:mcmc}
In the context of Bayesian uncertainty quantification, we are interested in generating samples from the
following probability distribution,
\begin{equation} \label{eq:bayes} p(\theta|y) \propto p(y|\theta) p(\theta),
\end{equation}
\noindent where $p(y|\theta)$ is the likelihood of the parameter set $\theta$ given observed data
$y_0,\ldots,\allowbreak y_{n-1}$, and $p(\theta)$ is the prior distribution of the parameters.  In many
applications the likelihood $p(y|\theta)$ is based on some physical model where we cannot derive a closed form expression for the
posterior density $p(\theta|y)$.  Markov Chain Monte Carlo (MCMC) has emerged over the last 30 years as one of
the most generally applicable and widely used framework for generating samples from the posterior distribution
\citep{gelman2013bayesian,green2015bayesian}.  The software used in this paper makes use of two MCMC
algorithms: the No U-Turn Sampler (NUTS) \citep{hoffman2014no} and the simplified manifold Metropolis Adjusted
Langevin Algorithm (smMALA) \citep{girolami2011riemann}.  NUTS is called via the Stan environment
\citep{carpenter2017stan}.  It is a variant of Hamiltonian Monte Carlo (HMC), which uses the gradient (first
derivative) of the posterior density, whereas smMALA uses the gradient and Hessian (first and second derivatives) of
the posterior density, smMALA is described in Algorithm \ref{alg:smmala}.

The error in estimates obtained from MCMC is approximately $C / \sqrt{N}$, where $N$ is the number of MCMC
iterations and $C$ is some problem-dependent constant.  In general it is not possible to demonstrate that MCMC
has converged, but there are several diagnostics that can indicate non-convergence, see Section 11.4 of
\citep{gelman2013bayesian} for more detail.  Briefly, there are two phases of MCMC sampling: burn-in and the
stationary phase.  Burn-in is finished when we are in the region of the parameter space containing the true
parameters.  In this paper we restrict ourselves to synthetic data examples.  In this case it is
straight-forward to assess whether the sampler is burnt in by checking whether the true parameters used to
simulate the data are contained in the credible intervals obtained from the generated samples.  In real data
applications, it is good practice to test MCMC sampling on a synthetic data problem that is analogous to the
real data problem.  During the stationary phase we assess convergence rate using the effective sample size,
\textit{N Eff}.  If we were able to generate independent samples from the posterior then the constant $C$ is
$\mathcal{O}(1)$.  MCMC methods generate correlated samples, in which case $C$ may be $\gg 1$.  A small
\textit{N Eff}  (relative to $N$) is an indicator of slow convergence.

If we are sampling from a multivariate target distribution, \textit{N Eff} for the $i$th component is equal to,
\begin{equation} \frac{S}{1 + 2 \sum_k \hat{\rho}_i(k)},
\end{equation}

\noindent where $S$ is the number of samples obtained during the stationary period, and $\hat{\rho}_i(k)$, is
an estimate of the autocorrelation at lag $k$ for the $i$th component of the samples.  This expression can be
derived from writing down the variance of the average of a correlated sequence (Chapter 11 of
\citep{gelman2013bayesian}).  The key point to note is that if the autocorrelation decays slowly \textit{N Eff} will
be relatively small.
\begin{algorithm}[ht]
\KwIn{Data, $y$; Initial value for $\theta = (\theta_1,\ldots,\theta_N)$;
\newline
Likelihood $l(y|\theta)$; Prior distribution $p(\theta)$.}
\Parameter{Step size, $h$; Number of iterations, $I$}
\For{$i = 2,\ldots,I$ }{
    Evaluate gradient and Hessian of the unnormalized log posterior,\newline
    \Indp $g_{\theta} = \nabla \bigg[\log[ \ l(y|\theta)\ p(\theta) \ ] \bigg]$ and $\mathbf{H}_{\theta}$\;
    Set $C = h^2 \mathbf{H}_{\theta}^{-1}$, and $m = \theta + \frac{1}{2} C\, g_{\theta}$\;
    Propose $\theta^* \sim q(\theta^*|\theta)=N(m, C)$\;
    Evaluate acceptance ratio, $\alpha(\theta, \theta^*) =
        \mathrm{min}\left[1, \dfrac{l(y|\theta^*) p(\theta^*) q(\theta|\theta^*)
}{l(y|\theta) p(\theta) q(\theta^*|\theta)} \right]$\;
    Draw $u \sim \mathrm{Uniform}(0,1)$\;
    \lIf{$u < \alpha(\theta, \theta^*)$}{Set $\theta = \theta^*$ }
}
\caption{smMALA}\label{alg:smmala}
\end{algorithm}

\subsection{Derivative computation}
\label{sec:deriv}
A central finite difference approximation to the first and second derivatives of the function $F(\theta): \mathbb{R}^N \rightarrow
\mathbb{R}^M$ can be computed as,
\begin{equation}
  \label{eq:d1}
  \frac{\partial F}{\partial \theta_i}  \approx \frac{F(\theta+h e_i) - F(\theta-h e_i)}{2 h}, \qquad
  \frac{\partial^2 F}{\partial \theta_i \partial \theta_j}  \approx
  \frac{\dfrac{\partial F}{\partial \theta_i}\big(\theta + h e_j\big) -
  \dfrac{\partial F}{\partial \theta_i}\big(\theta-h e_i\big)}{2 h}, \nonumber
\end{equation}

\noindent where $e_i$ is the $i$th Cartesian basis vector, and $h$ is a user-defined step-size.  For
first-order derivatives the default value we use is $\sqrt{\epsilon}|\theta_i|$, where $\epsilon$ is machine
epsilon for double types, i.e., if $\theta_i$is $\mathcal{O}(1)$, $h\approx10^{-8}$.  For second derivatives
the default value is $\epsilon^{1/3}|\theta_i|$, so that $h\approx 5\cdot 10^{-6}$.  More details on the
optimal step size in finite difference approximations can be found in \citep{nagblog-fd}.  First derivatives
computed using finite differences require $\mathcal{O}(N)$ function evaluations, where $N$ is the number of
input variables, and the optimal relative accuracy is around $10^{-6}-10^{-8}$ for double precision variables.  Second
derivatives require $\mathcal{O}(N^2)$ function evaluations, and the optimal accuracy is around $10^{-2}-10^{-4}$.

Derivatives that are accurate to machine precision ($10^{-14}-10^{-16}$ for double precision variables) can be
computed using AD tools.  AD tools generally use one of two modes: tangent (forward) or adjoint (reverse).
For a function with $N$ inputs and $M$ outputs, tangent mode requires $\mathcal{O}(N)$ function evaluations
and adjoint mode requires $\mathcal{O}(M)$ function evaluations.  In statistical applications our output is
often a scalar probability density, so adjoint AD will scale better with $N$ than either tangent mode or
finite differences in terms of total computation time.  Adjoint AD is implemented as follows using the Stan
Math Library, \citep{carpenter2015stan}.
\begin{verbatim}
// Define and init matrix with Stan's adjoint type
Matrix<stan::math::var,Dynamic,1> theta = input_values;

// Compute primal and gradient
stan::math::var lp = computation(theta);
lp.grad();

// Extract gradient
Matrix<double,Dynamic,1> grad_vec(theta.size());
for(int j=0; j<theta.size(); j++) grad_vec(j) = theta(j).adj();
\end{verbatim}

In the code above the function is evaluated using the \texttt{stan::math:var} scalar type rather than
\texttt{double}.  During function evaluation, derivative information is recorded.  When the \texttt{grad()}
function is called this derivative information is interpreted and derivative information is then accessed by
calling \texttt{adj()} on the input variables.  Similarly to the Stan Math Library, derivatives with
machine-precision accuracy can be computed using the NAG \dcocpp{} tool developed in collaborations with RWTH
Aachen University's STCE group, \citep{AIB-2016-08}.
\begin{verbatim}
// Define and init matrix with dco's adjoint type
typedef dco::ga1s<double> DCO_MODE;
typedef DCO_MODE::type Scalar;
Matrix<Scalar,Dynamic,1> theta = input_values;

// Enable dynamic activity analysis
for(int j=0; j<theta.size(); j++)
  DCO_MODE::global_tape->register_variable(theta(j));

// Compute primal and gradient
Scalar lp = computation(theta);
dco::derivative(lp) = 1.0;
DCO_MODE::global_tape->interpret_adjoint();

// Extract gradient
Matrix<double,Dynamic,1> grad_vec(theta.size());
for(int j=0; j<theta.size(); j++)
  grad_vec(j) = dco::derivative(theta(j));
\end{verbatim}

\dcocpp{} provides its adjoint type using the \texttt{dco::ga1s<T>::type} typedef, where \texttt{T}
is the corresponding primal type (e.g.\ \texttt{double}). Higher order adjoints can be achieved by
recursively nesting the adjoint type. Using this type, the program is first executed in the augmented
forward run, where derivative information is stored in the \texttt{global\_tape} data structure.
The \texttt{register\_variable} function initialises recording of the tape and facilitates dynamic
varied analysis~\citep{hascoet2005tbr}. Derivative information recorded during the augmented primal run is
then interpreted using the \texttt{interpret\_adjoint()} function and \texttt{dco::derivative} is used to
access the propagated adjoints.

\section{Results for noise driven harmonic oscillator}

\begin{figure}
  \centering
  \hspace{-2.2cm}
  \begin{minipage}[b]{0.60\linewidth}
    \centering
  \includegraphics[width=1.05\linewidth]{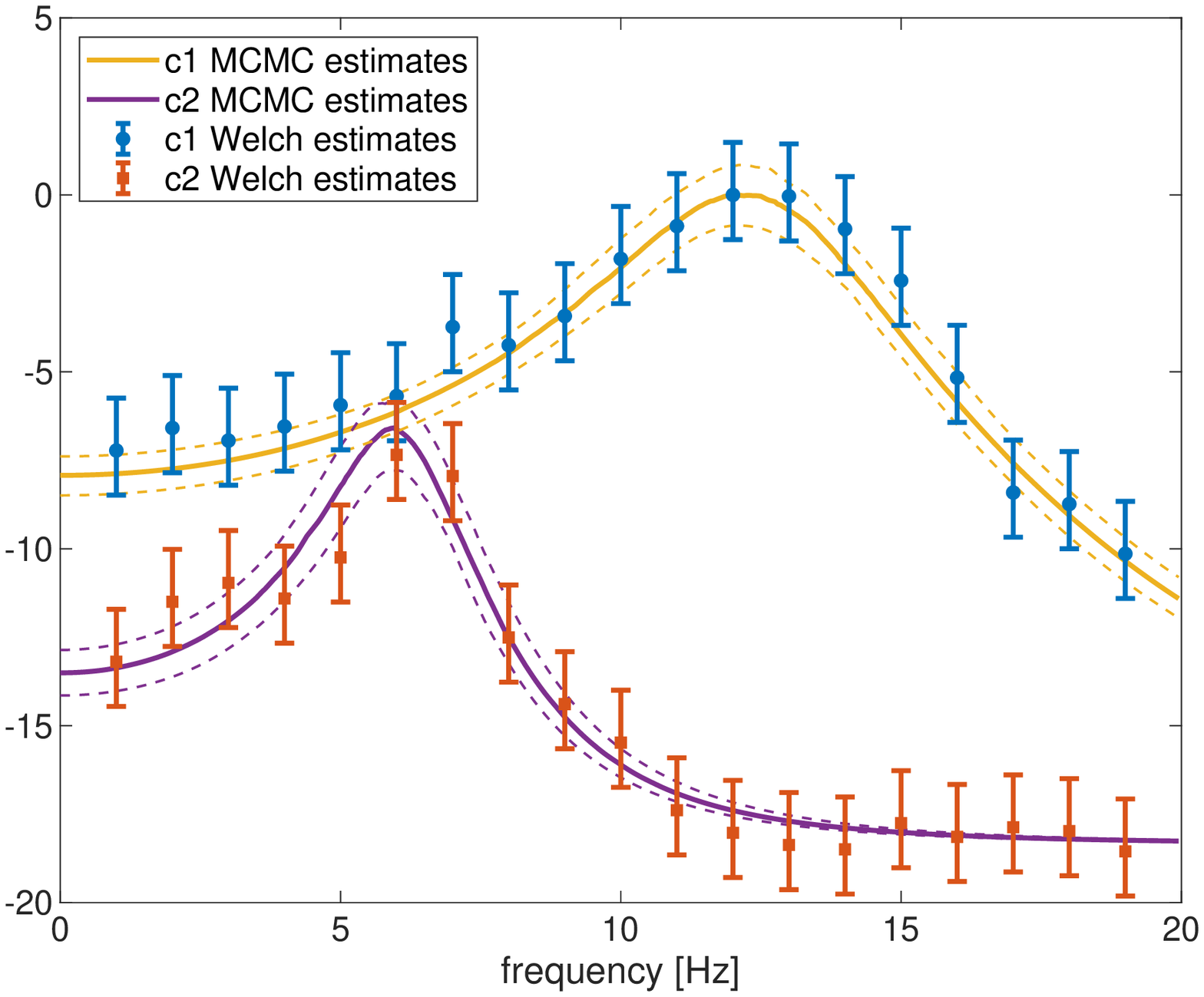}
  \end{minipage}%
  \begin{minipage}[b]{0.30\linewidth}
    \centering
  \begin{tabular}[b]{|c||c|c|c|c|}
    \hline
    \multirow{2}{*}  & \multirow{2}{*}{Actual} & \multicolumn{3}{c|}{Quantile} \\
    \hhline{|~||~|-|-|-|}
                     &  & 0.025 & 0.50 & 0.975 \\
    \hhline{|=||=|=|=|=|}
    $\omega_0(c_1)$ & 80 & 78.3 & 80.0 & 82.1 \\
    \hline
    $\omega_0(c_2)$ & 40 & 37.4 & 38.9 & 40.4 \\
    \hline
    $\sigma_{in}(c_1)$ & 100 & 94.0 & 101 & 109 \\
    \hline
    $\sigma_{in}(c_2)$ & 10 & 9.53 & 10.6 & 11.7 \\
    \hline
    $\zeta$ & 0.2 & 0.17 & 0.19 & 0.22 \\
    \hline
  \end{tabular}
  \vspace{1.5cm}
\end{minipage}
\caption{Spectral density and posterior parameter estimates for noise driven harmonic oscillator.  Synthetic data
  was generated by simulating from the model: duration = 20.0, time-step = 0.01.  Two datasets ($c_1$ and
  $c_2$) were generated with different parameter values for $\omega_0$ and $\sigma_{in}$ in each dataset
  (values are given in `actual' column).  The quantiles are estimated from a sequence of $10,000$ MCMC samples
  from the joint (posterior) distribution of all the parameters given all of the data.
  For example, 2.5\% of the MCMC samples had $\omega_0(c_1)$ values less than 78.3.}
\label{fig:ho_mcmc_spec}
\end{figure}

We now present the results of MCMC sampling using smMALA to estimate parameters of the noise driven
harmonic oscillator.  Then we demonstrate that, for this particular model, MCMC sampling can be accelerated by
using NUTS.  We also compare run times and sampling efficiency when AD is used to evaluate derivatives.  Here
we estimate parameters using synthetic data (i.e. data generated from the model).  This is a useful check that
the MCMC sampler is working correctly: we should be able to recover the parameters that were used to simulate
the data.  We generate a pair of datasets representing different conditions (which we label $c_1$ and $c_2$).
We assume that the $\omega_0$ and $\sigma_{in}$ parameters vary across conditions, but that $\zeta$ is fixed
across conditions, and that $\sigma_{obs}=0.05$ (i.e. it is known and does not need to be sampled).  In Table
\ref{fig:ho_mcmc_spec} we show that the $95\%$ credible intervals for each parameter include the actual
parameter values for all 5 parameters.  These results came from one MCMC run of $10,000$ iterations using the
Whittle likelihood and a uniform prior.  The step size parameter, $h$, was set to $1.0$.  There were no
burn-in iterations as the sampler was initialized at the true parameter values.  Figure \ref{fig:ho_mcmc_spec}
shows $95\%$ confidence intervals for the spectral density obtained using the Welch method alongside $95\%$
credible intervals for the spectral density estimated from $10,000$ MCMC iterations.  This is another useful
check: the spectral density predictions generated by sampled parameter sets are consistent with non-parametric
estimates of the spectral density.

Table \ref{tab:bench} uses the same model and the same datasets as above to compare smMALA with NUTS, and
finite differences with AD.  The AD implementation used in smMALA was dco's tangent over adjoint mode
(i.e. \texttt{dco::gt1s} combined with \texttt{dco::ga1s}).  In NUTS we used the dco's tangent mode for
computing the derivatives of the spectral density and Stan's adjoint mode (\texttt{stan::math::var}) for the
rest of the computation.  Given the most recent developments in the Stan Math Library (see release notes for
v3.2.0) it would likely be possible to use Stan's adjoint mode for the whole computation.  However this was
not the case when we started writing the code.  In general users may find that they need the more advanced
functionality of \dcocpp{} for part of their computation in order to obtain good performance.

The MCMC samplers were each run for 1,000 iterations.  The results were analyzed using Stan's
\texttt{stansummary} command-line tool.  To account for correlation between MCMC samples we use the \textit{N
Eff} diagnostic defined in Section \ref{sec:mcmc} to measure sampling efficiency, min \textit{N Eff}  is the
minimum over the 5 model parameters.  Table \ref{tab:bench} shows that, for the noise driven harmonic
oscillator, we can accelerate MCMC sampling by a factor of around 3-4 through the combined effect of more
efficient MCMC sampling (NUTS rather than smMALA) and more accurate derivative evaluations (AD rather than
finite differences).  In some cases the \textit{N Eff / s} is higher for finite differences than for AD,
because of the small number of input variables.  However, even for this simple model the NUTS min \textit{N
Eff} is higher with AD than with central finite differences suggesting that the extra accuracy in the
derivatives results in more efficient sampling per MCMC iteration.

In the context of Bayesian uncertainty quantification what we are interested in is whether the MCMC samples
are an accurate representation of the true posterior distribution.  A necessary condition for posterior
accuracy is that the true parameters are (on average) contained in the credible intervals derived from the
MCMC samples.  This is the case for all the different variants of MCMC that we tested.  The main differences
we found between the different variants was the sampling efficiency.  As discussed in Section \ref{sec:mcmc},
a sampling efficiency that is $3-4$ time greater means a reduction in the value of $C$ in the
Monte Carlo error ($C / \sqrt{N}$) by that factor.  Another way of interpreting this is that we could
reduce the number of MCMC iterations by a factor of $3-4$ and expect to obtain the same level of accuracy in
the estimated posterior distribution.
\begin{table}[!htb]
  \caption{Noise-driven harmonic oscillator benchmarking results on Intel 2.50GHz CPU, Ubuntu (18.04) Linux.}
  \begin{tabular}{|l|l|l|l|l|}
    \hline
    MCMC sampler & Derivative implementation & CPU time (s) & min \textit{N Eff} & min \textit{N Eff} / s \\
    \hline
    smMALA & Finite differences  & 6.2  & 150  & 24  \\
    smMALA & \dcocpp             & 8.2  & 150  & 18  \\
    NUTS   & Finite differences  & 8.5  & 384  & 46  \\
    NUTS   & Stan and \dcocpp    & 6.0  & 503  & 84 \\
    \hline
  \end{tabular}
  \label{tab:bench}
\end{table}

%\pagebreak
\section{Discussion}

The Whittle likelihood function is written to be polymorphic over classes derived from the
a base class called \texttt{Stable\_sde}.  The function signature includes a reference to the base class.
\begin{verbatim}
template<typename SCALAR_T>
SCALAR_T log_whittle_like(Stable_sde<SCALAR_T> & m, ...);
\end{verbatim}

Classes that are derived from the \texttt{Stable\_sde} base class must define the following pure virtual
function.
\begin{verbatim}
virtual Matrix<SCALAR_T, Dynamic, Dynamic> sde_jacobian(...) = 0;
\end{verbatim}

The spectral density evaluation (Equation \eqref{eq:spec_eq}) is implemented in the base class, reducing the
effort required to do spectral data analysis for other stable SDEs.  The smMALA sampler is written
to be polymorphic over classes derived from a base class called \texttt{Computation}.  Classes that are
derived from the \texttt{Computation} base class must define the following pure virtual function.
\begin{verbatim}
virtual SCALAR_T eval(Matrix<SCALAR_T,Dynamic,1>& theta,...) = 0;
\end{verbatim}

This function should evaluate the posterior density in the user's model.  The gradient and Hessian of the
posterior density are implemented in the \texttt{Compu\-tation} class.  This reduces the effort required to
use NUTS or smMALA for other computational models.  Classes derived from \texttt{Stable\_sde} or from
\texttt{Computation} can be instantiated with several different scalar types including \texttt{double},
\texttt{stan\allowbreak::math\allowbreak::var}, and a number of \dcocpp{} types.  This enables automatic
evaluation of first and second derivatives.  The gradient and Hessian functions need a template specialization
to be defined in the base class for each different scalar type that is instantiated.

We conclude with some comments regarding the choice of MCMC sampler and the method for evaluating derivatives.
Our software only implements derivative-based MCMC as this tends to be more computationally efficient than
other MCMC methods \citep{hoffman2014no,penny2016annealed}.  Derivative-based MCMC samplers can be further
subdivided into methods that use higher-order derivatives, such as smMALA (sometimes referred to as
Riemannian) and methods that only require first-order derivatives (such as NUTS).  Riemannian methods tend to
be more robust to complexity in the posterior distribution, but first-order methods tend to be more
computationally efficient for problems where the posterior geometry is relatively simple.  We recommend
using smMALA in the first instance, then NUTS as a method that may provide acceleration in MCMC sampling, in
terms of effective samples per second, \textit{N Eff /s}.  Regarding the derivative method, finite difference
methods often result in adequate performance for problems with moderate input dimension (e.g. 10-20), at least
with smMALA.  But for higher-dimensional problems (e.g. Partial Differential Equations or Monte Carlo
simulation) we recommend accelerating derivative computation using adjoint AD
\citep{nagsite-ad,naumann2018adjoint}.  The software presented enables users to implement and benchmark all
these alternatives so that the most appropriate methods for a given problem can be chosen.

\bibliographystyle{apalike}
\bibliography{refs}




% LNCS Homepage, \url{http://www.springer.com/lncs}. Last accessed 4
% \end{thebibliography}
\end{document}